\documentclass[twocolumn, tighten, times, astrosymb]{aastex631}

\newcommand\eri{Eri~{\sc II}}
\newcommand\msun{$M_{\odot}$}
\newcommand\hst{\textit{HST}}
\newcommand\jwst{\textit{JWST}}

\received{July 6, 2022}

\accepted{March 10, 2023}

\shorttitle{Eri~{\sc II} Globular Cluster}
\shortauthors{Weisz et al.}

\graphicspath{{./}{figures/}}

\begin{document}

\title{On the Reionization-Era Globular Cluster in Low-Mass Galaxy Eridanus {\sc II}}

\correspondingauthor{Daniel R. Weisz}
\email{dan.weisz@berkeley.edu}

\author[0000-0002-6442-6030]{Daniel R. Weisz}
\affiliation{Department of Astronomy\\ University of California \\ Berkeley, CA 94720, USA}

\author[0000-0002-1445-4877]{Alessandro Savino}
\affiliation{Department of Astronomy\\ University of California \\ Berkeley, CA 94720, USA}

\author[0000-0001-8416-4093]{Andrew E. Dolphin}
\affiliation{Raytheon Technologies, 1151 E. Hermans Road, Tucson, AZ 85756, USA}
\affiliation{Steward Observatory, University of Arizona, 933 N. Cherry Avenue, Tucson, AZ 85719, USA}

\begin{abstract}

Using color-magnitude diagrams from deep archival \textit{Hubble Space Telescope} imaging, we self-consistently measure the star formation history of Eridanus~{\sc II} (Eri~{\sc II}), the lowest-mass galaxy ($M_{\star}(z=0) \sim 10^5 M_{\odot}$) known to host a globular cluster (GC), and the age, mass, and metallicity of its GC.  The GC ($\sim13.2\pm0.4$~Gyr, $\langle$[Fe/H]$\rangle = -2.75\pm0.2$~dex) and field (mean age $\sim13.5\pm0.3$~Gyr, $\langle$[Fe/H]$\rangle = -2.6\pm0.15$~dex) have similar ages and metallicities.  Both are reionization-era relics that formed before the peak of cosmic star and GC formation ($z\sim2-4$). The ancient star formation properties of Eri II are not extreme and appear similar to $z=0$ dwarf galaxies.  We find that the GC was $\lesssim4$ times more massive at birth than today and was $\sim$10\% of the galaxy’s stellar mass at birth. At formation, we estimate that the progenitor of Eri~{\sc II} and its GC had $M_{\rm UV} \sim -7$ to $-12$, making it one of the most common type of galaxy in the early Universe, though it is fainter than direct detection limits, absent gravitational lensing. Archaeological studies of GCs in nearby low-mass galaxies may be the only way to constrain GC formation in such low-mass systems.  We discuss the strengths and limitations in comparing archaeological and high redshift studies of cluster formation, including challenges stemming from the Hubble Tension, which introduces uncertainties into the mapping between age and redshift.

\end{abstract}


\section{Introduction} \label{sec:intro}

Metal-poor globular clusters (GCs) are likely relics of the early Universe.  They are thought to have formed in low-mass halos at high-redshifts and may represent an extreme mode of star formation that is non-existant or rare at low-redshifts \citep[e.g.,][]{peebles1968, fall1985, elmegreen1997}.  

One obstacle to deciphering their origins is a paucity of suitable constraints on both metal-poor GCs and their host galaxies.  The sensitivity and resolving power of the \textit{Hubble Space Telescope} (\hst) and the \textit{James Webb Space Telescope} (\jwst) have revealed a number of possible young clusters and/or compact star forming clumps that may be progenitors of the GCs found in the local Universe \citep[e.g.,][]{vanzella2017, vanzella2019, zick2020, vanzella2022, pascale2023, vanzella2023, welch2023}.  However, given the high stellar masses and star formation rates (SFRs) associated with these objects, they are likely one of the most massive and extreme examples of cluster formation in the early Universe, and are unlikely to sample formation of typical metal-poor GCs.  Moreover, their host galaxies are likely progenitors of MW-like systems at $z=0$ rather that the very low-mass systems thought to host metal-poor GCs \citep[e.g.,][]{kravtsov2005}.  Due to their faintness, direct detection and characterization of the the lower mass GCs and their host galaxies is likely to prove challenging even with the sensitivity of \jwst.

At low redshift, we have access to a larger population of metal-poor GCs, which can be studied in great detail.  However, the majority of them reside in massive galaxies such as the Milky Way (MW) where the process of hierachical accretion of GCs, and in some cases their disruption into streams, obfuscates the connection between metal-poor GCs and their birth galaxies \citep[e.g.,][]{kravtsov2005, zaritksy2016, choksi2018, el-badry2019, kruijssen2019a, garcia2022, chen2023}. 

Historically, only a handful of GCs are known reside in nearby dwarf galaxies (e.g., \citealt{humason1956, ables1977, hoessel1982, hodge1999, grebel2000, cook2012, caldwell2017, cole2017, forbes2018, beasley2019, larsen2020, pace2021, larsen2022}), though new search efforts are rapidly increasing this number \citep[e.g.,][]{minniti2021, carlsten2022}.  The sample of GCs in low-mass galaxies has provided rich insight into the formation and evolution of metal-poor GCs in their birth environment. For example, a series of spectroscopic studies have shown that GCs in moderate-mass dwarf galaxies can only have been a factor of a few times more massive in the past, providing stringent constraints on the mass budget problem in these environments \citep[e.g.,][]{larsen2012, larsen2014}.  Similarly, the locations of the GCs within low-mass galaxies can be used to constrain the cluster formation mechanisms \citep[e.g.,][]{zaritsky2022}, while basic properties such as age and metallicity can be used to constrain a wide variety of astrophysics from the contribution of GCs to reionization to the halo masses and accretion histories of the host galaxies.

The Local Group (LG) ultra-faint dwarf galaxy (UFD) Eridanus~{\sc II} (\eri; \citealt{bechtol2015, koposov2015}) provides a unique opportunity to simultaneously study a metal-poor GC and its birth galaxy in the lowest mass galaxy known that hosts a GC \citep[e.g.,][]{forbes2018}. \eri\ has deep \hst\ imaging of its resolved stellar population enabling an archaeological characterization of the galaxy and GC in the early Universe.  Prior \hst-based studies have shown \eri\ to be an ancient ($>13$~Gyr old) and metal-poor ([Fe/H] $\lesssim -2.3$) galaxy \citep[e.g.,][]{gallart2021, simon2021, fu2022}.  However, there has been less focus on the GC and its relationship to the host galaxy and cluster formation more broadly.

In this paper, we self-consistently characterize the stellar populations of \eri's GC and field populations from color-magnitude diagram (CMD) analysis and combine our measurements with simple assumptions to constrain formation scenarios for a metal-poor GC in an extremely low-mass galaxy. We describe our photometric reduction in \S \ref{sec:data} and summarize our analysis methodology in \S \ref{sec:methods}.  We present our results and place them into a broader context in \S \ref{sec:results}.

\section{Data \& the Color-Magnitude Diagrams} \label{sec:data}

We use archival HST/ACS F475W, F606W, and F814W imaging that was taken as part of two separate HST programs (GO-14224 and GO-14234).  The deep imaging from both programs was designed to extend well-below the ancient main sequence turnoff (MSTO), enabling a `gold standard' reconstruction of \eri's properties at all cosmic epochs.

We performed point spread function (PSF) photometry on each \texttt{flc} image using \texttt{DOLPHOT}, a widely-used PSF fitting package that has an HST/ACS specific module \citep{dolphin2000b, dolphin2016}.  We used the \texttt{DOLPHOT} parameters and culling criteria specified in \citet{williams2014}, which were optimized for the disk of M31 and are well-suited for the comparatively less crowded field and GC populations of \eri.

We characterized uncertainties in the photometry of the GC and field population separately. Following resolved cluster analysis in \citet{johnson2016}, we generate $50,000$ artificial star tests (ASTs) that trace the light profile of the GC \citep[][]{simon2021}. For the field population, we uniformly distributed $500,000$ ASTs throughout the non-cluster region of the HST images.  The GC and field photometry is 100\% complete $>1$~mag below the oldest MSTO.

\begin{figure}[t!]
\epsscale{1.15}
\plotone{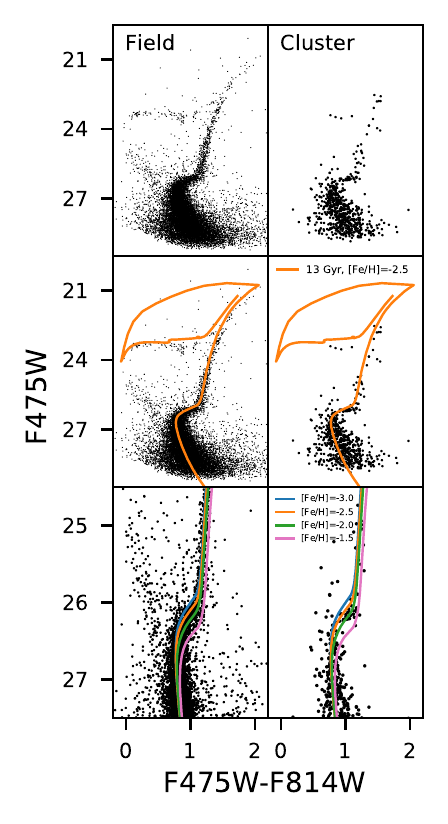}
\caption{The HST/ACS F475W-F814W CMDs of \eri\ and its cluster.  The GC CMD contains stars within $2r_h$ (half-light radius) and it has been excluded from the field CMD on the left.  The middle panels show a BaSTI isochrone at our measured HB distance.  The bottom panels are zoomed in on the MSTO regions use in our fitting. Over-plotted are varying [Fe/H] BaSTI models at a fixed age of 13.5~Gyr.  The cluster and field population appear qualitatively similar in age and metallicity. \label{fig:cmds}}
\end{figure}

Figure \ref{fig:cmds} shows the F475W-F814W CMDs of the field population and GC.  The GC CMD includes stars within $2r_h$ and field population CMD is all stars outside this radius.  Our field CMD is visually similar to those presented and discussed in \citet{gallart2021}, while our cluster CMD is similar to that presented in \citet{martinezvazquez2021}.  The overlaid BaSTI isochrones \citep{hidalgo2018} qualitatively show that \eri\ is predominantly an ancient, metal-poor population as found in other studies.  The presence of a blue and red horizontal branch may indicate a population of mixed age and/or broad metallicities, the latter of which is also reported in the literature \citep[e.g.,][]{li2017, fu2022}, or could reflect physics such as variable mass loss or helium abundance \citep[e.g.,][]{milone2017, savino2019}.  There is a clear blue plume (i.e., a vertical sequence of stars extending above the ancient MSTO), which may indicate a younger population and/or blue stragglers. Both \citet{gallart2021} and \citet{simon2021} show that the number of stars in this region of the CMD are consistent with expectations for blue stragglers.

\eri's GC also appears to be ancient and metal-poor.  \citet{simon2021} and \citet{martinezvazquez2021} comment on the visual similarity between the field and cluster populations.  The former study notes there may be a small offset in the colors of the GC and field populations near the sub-giant branch.  From the bottom panel of Figure \ref{fig:cmds}, it appears that the GC may be slightly more metal-poor than the field population; formal CMD fitting, i.e., a goal of this paper, is required to quantify such subtle differences.

\section{Methodology} \label{sec:methods}

\subsection{Distance}

Literature distances to \eri\ vary from $\mu=22.60$ to $22.90$ \citep[e.g.,][]{bechtol2015, koposov2015, simon2021, gallart2021, martinezvazquez2021, nagarajan2021}. Three of these distances are based on RR Lyrae, which are considered `gold standards' for Pop {\sc II} stars.  However, they are in modest tension: $\mu=24.65$ \citep{simon2021}, $\mu=24.67$ \citep{nagarajan2021}, and $\mu=24.84$ \citep{martinezvazquez2021}, making it challenging to know which one to use.  These discrepancies could be due to a variety of factors (e.g., light curve sampling, analysis techniques, zero point differences). Resolving these tensions are beyond the scope of this paper. For simplicity, we instead derive our own distance to \eri\ using the luminosity of the HB following \citet{mcquinn2015} and \citet{weisz2019a}.

Briefly, we convert F606W to a pseudo V-band using the filter transformations provided in \citet{saha2011}. F606W is closer to V-band than F475W making it a more suitable choice. We fit the pseudo V-band luminosity function of the region around the HB using a mixture of a Gaussian and exponential function.  We find the most likely pseudo V-band HB magnitude to be $m_V = 23.30$, which yields $\mu=22.79\pm0.1$ for $A_{\rm V}=0.038$~mag \citep{schlafly2011} and $M_{\rm V,0,HB} = +0.47$ \citep{weisz2019a}. This distance is consistent with most values reported in the literature, including falling within 1-1.5$\sigma$ of all of the RR Lyrae distances.  Isochrones at this distance and extinction show a good match to the observed CMDs (Figure \ref{fig:cmds}).  We adopt $\mu=22.79\pm0.1$ for the remainder of this analysis.

\subsection{Modeling the Field and Cluster Color-Magnitude Diagrams}

We measure the SFH of the field population and the age, mass, and metallicity of the GC using \texttt{MATCH} \citep{dolphin2002} and the F475W-F814W CMD, for which the broad color baseline is particularly sensitive to age and metallicity at the main sequence turnoff \citep[e.g.,][]{cole2007, monelli2010a, gallart2015}. \texttt{MATCH} is a widely used package for modeling CMDs of nearby galaxies and star clusters \citep{weisz2014a, weisz2016, johnson2016, skillman2017}.  We follow implementation details described in these papers.

Briefly, for a given set of stellar evolution models, a specified initial mass function (IMF), distance, extinction, and binary fraction, \texttt{MATCH} generates a set of synthetic simple stellar populations (SSPs) over a range of ages and metallicities. These SSPs are linearly combined to form a model composite CMD. The CMD is partitioned into small color and magnitude bins (i.e,. a Hess diagram).  The composite CMD is convolved with the error distribution from the ASTs and a contamination model CMD (e.g., foreground stars) is linearly added.  The resulting synthetic CMD is compared to the observed CMD using a Poisson likelihood function.  \texttt{MATCH} repeats this process either as a full grid search or by a maximum likelihood optimization.

\begin{figure}[ht!]
\plotone{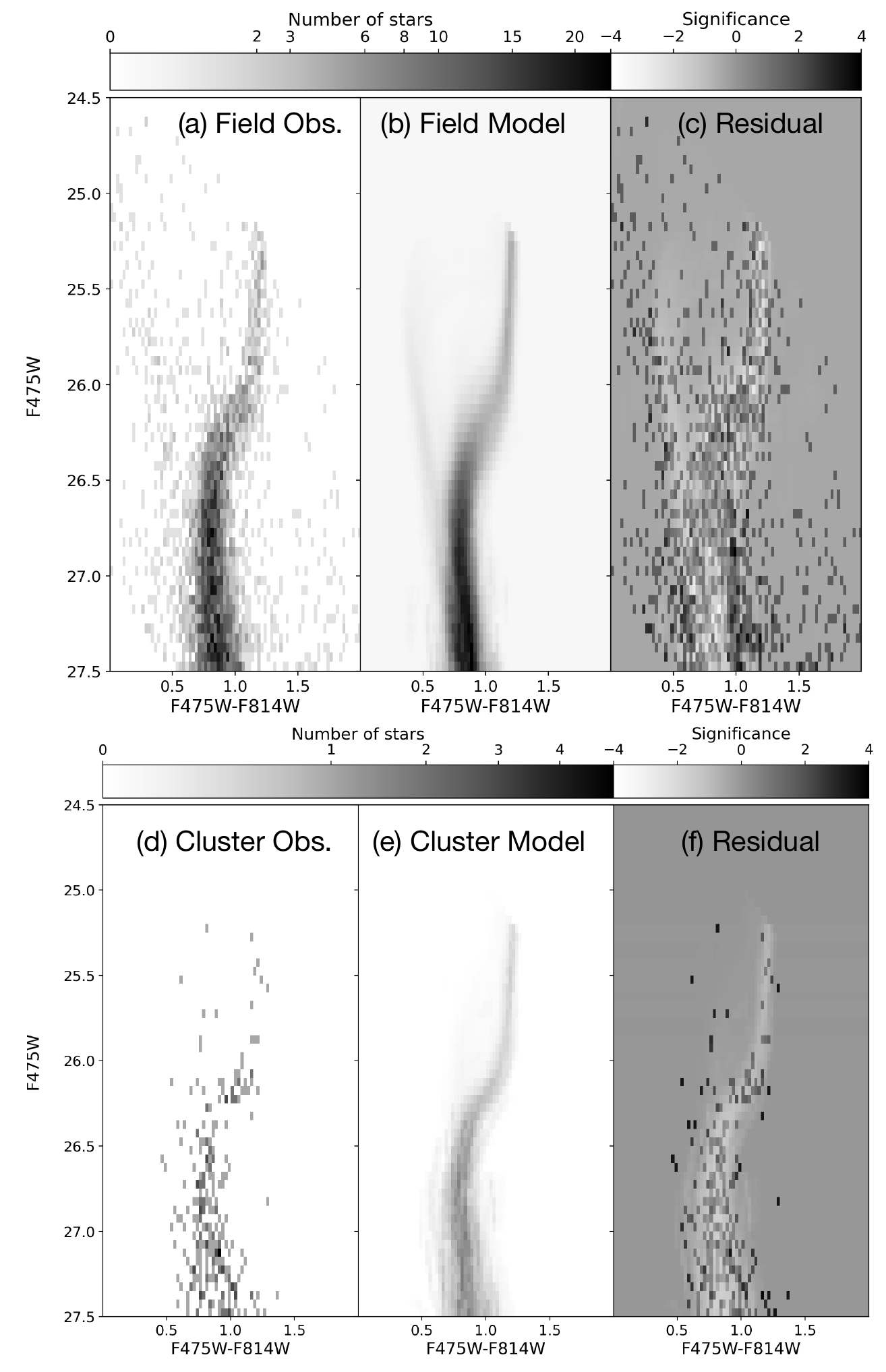}
\caption{Fit quality for \eri\ (top row) and its GC (bottom row). Plotted are the Hess diagrams for the data (left), the best fit models (middle), and residual normalized to $\sigma$ (right). The best fit models for both field and cluster population show no significant and/or systematics and overall appear to be good fits to the data.} \label{fig:hess}
\end{figure}

For this analysis, we adopt a Kroupa IMF \citep{kroupa2001} from 0.08 to 120 \msun, $\mu=22.79$~mag, $A_{\rm V} = 0.038$ \citep{schlafly2011}, and a binary fraction of 0.35 with a uniform mass ratio distribution.  We use the BaSTI \citep{hidalgo2018} stellar models, as they include heavy element diffusion and their publicly available models encompass the low stellar metallicites ([Fe/H] $<-3$) found in \eri\ \citep[][]{li2017, fu2022}.  We measured the SFH in two iterations.  First, we generate a grid of SSPs with coarse time ($\log(\Delta t) = 0.05$~dex) and metallicity ($\log(\Delta [M/H]) = 0.1$~dex)  resolutions over the ranges $10.15 \le \log(t) \le 9.0$ and $-3.5 \le \log([M/H]) \le -1.5$.  This allowed us to get a broad sense of the SFH and evaluate the significance of the `blue plume' population.  We found that the majority of field star formation occurred in the oldest time bin, while $\sim4$\% of the stellar mass formed 2-3~Gyr ago.  This younger population appears consistent with blue straggler contributions and is consistent with the findings of other analyses of \eri.  Given the high SNR of the \eri\ data, we opted to measure the SFH at much higher time resolution.  In this case, we generate a grid of SSPs with fine time ($\log(\Delta t) = 0.01$~dex) and metallicity ($\log(\Delta [M/H]) = 0.05$~dex) over the ranges $10.14 \le \log(t) \le 9.8$ and $-3.5 \le \log(t) \le -1.5$. The lower limit on age was selected due to make for reasonable computational efficiency, even though it formally excludes the blue straggler contribution.  This has no impact on any of our results. We restrict our analysis to the region of the field and GC CMDs around the MSTO (see Figure \ref{fig:hess}).

For the field SFH, we exclude stars within $2 r_h$ of the GC.  We include a model for Galactic foreground contamination \citep{dejong2010b}, even though only a few $\sim5$ MW interlopers are expected in this field of view and to have passed all photometric quality cuts. We place a prior on the age-metallicity relationship such that it must increase monotonically, with an allowed dispersion of 0.4~dex, which is based on RGB star metallicity studies of \eri. Given all these parameters, \texttt{MATCH} finds the maximum likelihood solution. To compute random uncertainties, we use a Hamiltonian Monte Carlo \citep[HMC;]{duane1987} approach described in \citet{dolphin2013} and implemented in \citet{weisz2014a}.  We initialize the HMC chains at the maximum likelihood point and draw 5000 samples after burn-in.  Random uncertainties are the narrowest 68\% interval around the best fit SFH.

We model the GC using only stars located within $2r_h$.  We use the field population from $>2r_h$ as a background CMD and allow \texttt{MATCH} to linearly scale its contribution while solving for the GC parameters.  This has proven to be an effective decontamination method in practice \citep[e.g., for clusters in the disk of M31;][]{weisz2015, johnson2016}. We assume the cluster is an SSP and conduct a full grid search over the entire age and metallicity grid defined above. Assuming flat priors in age and metallicty, we compute the marginalized distributions for age and metallicty of the GC.  For uncertainties, we report the best fit age of each GC and the 68\% confidence around the best fit and add the grid resolution (0.01~dex in age, 0.05 dex in metallicity) in quadrature to account for the finite width of the basis functions \citep[e.g.,][]{weisz2016}. 

Figure \ref{fig:hess} shows the observed, best fit model, and residual Hess diagrams for the field and GC CMDs.  For both the field and GC, the residuals show that age and metallicity sensitive regions of the observed CMD, i.e., the MSTO and SGB, are matched by the model at the $<2$-$\sigma$ level.  The highest areas of disagreement are a handful of pixels below the oldest MSTO that is not age sensitive.  Overall, the observed CMDs are well-described by the best-fit models.

\begin{figure}[t!]
\plotone{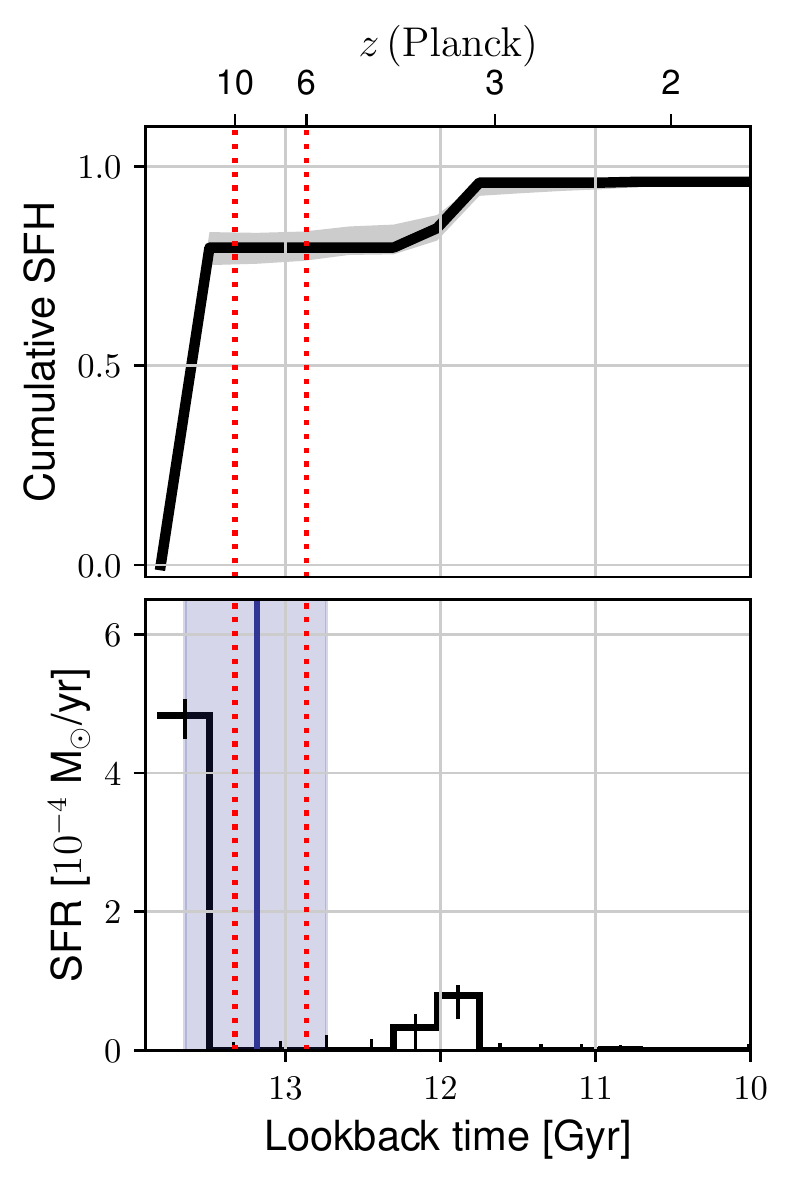}
\caption{The cumulative (top) and differential (bottom) SFHs of \eri.  The black lines are the best fit SFHs and the uncertainties reflect the 68\% confidence intervals.  The dark blue line shows the best fit GC age for each model and the light blue shaded region is the 68\% confidence interval.  The red dashed lines indicates the timing of reionization assuming a \citet{planck2020} cosmology. }  
\label{fig:sfhs}
\end{figure}

\section{Results} \label{sec:results}

Figure \ref{fig:sfhs} shows the SFH of \eri's field population for the BaSTI and MIST models with the corresponding GC age over-plotted.  Uncertainties for the field SFH and GC are the 68\% confidence intervals. We list the measured properties of the GC in Table \ref{tab:gc}.

The field population of \eri\ is ancient: $\sim80$\% of the stellar mass formed prior to 13.5~Gyr ago ($z\sim10$), which is nearly identical to field SFHs for \eri\ in the literature \citep[][]{alzate2021, gallart2021, simon2021}.  We find a small secondary burst of star formation $\sim11-12$~Gyr ago, but it is of low significance. This secondary event could contribute to \eri's broad HB morphology, though the metallicity spread in \eri\ may also contribute. The mass-weighted age of \eri's field population is 13.5$\pm0.3$~Gyr. As with many UFDs in the MW halo \citep[e.g.,][]{brown2014, weisz2014a, sacchi2021}, \eri\ appears to be a reionization-era galaxy within the \textit{Planck}-based $\Lambda$CDM cosmology. 

We find the mean metallicity of the field population to be $\langle$[Fe/H]$\rangle= -2.6\pm0.15$~dex, which is similar to RGB star metallicities \citep[e.g.,][]{li2017, fu2022}.  If \eri\ is $\alpha$-enhanced, the values of [Fe/H] may be a few tenths of a dex lower; BaSTI $\alpha$-enhanced models were not available at the time of our analysis, however.  

We measure the stellar mass within the ACS field to be $M_{\star}\sim1.9_{-0.03}^{+0.04}\times10^5$ \msun. Following \citet{gallart2021}, we multiply a factor of 2 to account for the fact that the ACS field encompasses $\sim50$\% of the galaxy's integrated optical light get a total stellar mass of $M_{\star}\approx3.8\times10^5$ \msun.  This is the stellar \textit{birth} mass of the galaxy as the CMD fitting code uses stellar birth masses and does not account for mass loss due to stellar evolution.  We discuss the present day stellar mass in \S \ref{sec:masstolight}.

\eri's GC is ancient, metal-poor, and has a modest stellar mass (Table \ref{tab:gc}). We find a GC metallicity of $\langle$[Fe/H]$\rangle= -2.75\pm0.2$~dex, which is marginally more metal-poor than the field population, though consistent within uncertainties.  The best fit cluster age, $13.2\pm0.4$~Gyr, is slightly younger than the field, though formally consistent ages within 1-$\sigma$ uncertainties.

As a check on robustness of our modeling, we also fit the field population and GC using the MIST stellar libraries \citep{choi2016}.  The MIST-based field SFH is nearly identical ($\sim90$\% of the stellar mass formed prior to $\sim$13.5~Gyr ago) to our solution using the BaSTI models.  Similarly, the MIST-based age of the GC is 13.2$\pm0.4$~Gyr, which is the same as our BaSTI fit.  The MIST results are systematically several tenths of a dex more metal-rich.  The similarity of SFHs but differences in recovered metallicities when using different stellar libraries has previously been identified and quantified in more detail (e.g., see the Appendix in \citealt{skillman2017}). 
In both cases, the formation epoch of the GC is clearly older than than the peak GC formation epoch ($z\sim2-4$) predicted by several recent models of GC formation \citep[e.g.,][]{choksi2018, el-badry2019, reinacampos2019}; though some models find the peak of metal-poor GC formation to be at $z\gtrsim5$ \citep[e.g.,][]{renzini2017, boylankolchin2018}.

We measure the stellar mass of the GC to be $M_{\star} \approx 8.6\times10^3$\msun.  This mass is based on CMD modeling, which models the observed star counts at the present-day and sums their \textit{birth} masses to get a total stellar mass of the population.  That is the CMD mass is not the present-day mass because it uses initial masses from the stellar models and it is not the birth mass because it does not account of dynamical mass loss effects.  However, with some modest assumptions we can translate the CMD-based mass into the present-day and birth mass of the cluster.  We calculate these quantities in the next section.

\begin{table}[]
\begin{tabular}{lcc}
\hline \hline
            Quantity                    & Field        & GC          \\
\hline \hline
Age (Gyr)                                 & $13.5\pm0.3$ & $13.2\pm0.4$ \\ 
$\langle${[}Fe/H{]}$\rangle$ (dex) & $-2.6\pm0.15$ & $-2.75\pm0.2$ \\
$M_{\star, CMD}$ ($10^3 M_{\odot}$)            & 380 & 8.6 \\ 
$M_{\star, Birth}$ ($10^3 M_{\odot}$)            & 380 & $\lesssim34$ \\ 
$M_{\star, z=0}$ ($10^3 M_{\odot}$)            & 150 &  5.2  \\
$M_{V}/L_{V}(z=0)$ &     $\approx2.5$       & $\approx2.5$  \\
$\dot{M}_{\star}$ ($10^{-3}$ \msun\ yr$^{-1})$ &     $1-7$       & $1-5$  \\

$\Sigma_{\rm SFR}$ (\msun\ yr$^{-1}$ kpc$^{-2}$) &    $0.001-0.9$        & $5-500$\\
$M_{\rm UV, formation}$ (mag) &     $\sim -9\pm2.5$       & $\sim -12.5$ \\

\hline
\end{tabular}

\caption{Measured or estimated properties of the GC in \eri.  We list three types of stellar mass for the cluster: the stellar mass measured directly from the CMD ($M_{\star, CMD}$); the stellar mass at GC birth $M_{\star, Birth}$, which includes a correction for mass loss due to dynamical effects; the present-day stellar mass $M_{\star, z=0}$, which includes dynamical mass loss and mass loss due to stellar evolution.  We discuss how each quantity is estimated in \S \ref{sec:results}.}
\label{tab:gc}
\end{table}

\section{Discussion}

\subsection{Cluster Birth Mass and Constraints on Dynamical Mass Loss}

The birth masses and amount of dynamical mass loss in GC are key to several open questions including the origin of multiple populations, the contribution of GCs to reionization, and their long-term survival \citep[e.g.,][]{schaerer2011, conroy2012a, bastian2015, kruijssen2015, renzini2015, bastian2018, boylan-kolchin2018, krumholz2019}.

However, the mass lost over the lifetime of a GC due to dynamical effects is challenging to measure resulting in a paucity of empirical constraints (see discussion in \citet{bastian2018} and references therein), particularly in in low-mass galaxies.  One of the best constraints in a low-mass system comes from \citet{larsen2012} which estimates dynamical mass loss from the GCs in Fornax.  They compare the number and mass of metal-poor stars in the field of Fornax to its GC population to find that the GCs were no more than 3-5 times more massive at birth compared to the present day. Using the same technique, \citet{larsen2014} find comparable amounts of mass loss in slightly more massive dwarf galaxies WLM and IKN.

We use a similar approach to estimate the amount of mass lost by the GC in \eri.  Our CMD modeling provides the total mass of field stars as a function of metallicity, i.e., a mass-weighted metallicity distribution function (MDF).  Within the 1-$\sigma$ range of the cluster metallicity, i.e.,  [Fe/H]$=-2.55$ to $-2.95$~dex, we find the cluster-to-galaxy stellar mass ratio to $\sim3.5\pm0.5$.  This estimate includes the areal correction applied to the field population. This is an upper limit on the amount of stellar mass lost by the GC through dynamical effects over its lifetime , i.e., if the GC was entirely responsible for field stars in this metallicity range then it would have been 3-4 times more massive at birth. 

We multiply the CMD-based mass by 4 and and find a maximum birth mass of $M_{\star, Birth} \lesssim 3.4\pm0.4\times10^4$ \msun.  At birth, the GC was $\sim10$\% of the galaxy stellar mass.

Our upper limit on dynamical mass loss for \eri's GC is comparable to the upper limit of 3-5 reported by \citet{larsen2012} for Fornax, WM, and IKN. The consistency between mass loss in \eri\ and Fornax (the closest to \eri\ in stellar mass) is interesting as Fornax has 100 times more stellar mass than \eri\ at the present day, which would suggest a stronger disruption mechanisms \citep[e.g.,][]{kruijssen2015}. However, because this method only provides for upper limits, it is not possible to identify subtle differences in mass loss as a function of galaxy mass.  Nevertheless, there are now two data points in the low-mass galaxy regime that suggest dynamical mass loss is no more than a factor of a few.

A moderate birth mass, and moderate mass loss, for \eri's GC is broadly in line with some theoretical predictions for GC formation in low-mass halos \citep[e.g.,][]{kruijssen2015, choksi2018, kruijssen2019}.  In massive galaxies like the MW, lower mass GCs ($M_{\star, birth} < 10^5$) would be unlikely to survive for a Hubble time in the harsh galactic environment (e.g., due to tidal shocks, collisions with gas clouds, migration, the density of the host galaxy/halo).  However, these effects are thought to be much weaker in a low-mass galaxy like \eri, which enables the long-term survival of moderate mass GCs.  The long lifetime of \eri's GC and modest dynamical mass loss support this broad picture.

\subsection{Connections to Select Theoretical Models of Cluster Formation in Low-Mass Halos}

The GC in \eri\ provides a useful data point for constraining various theoretical models of GC formation in the low-mass galaxy regime.  For example, \citet{kruijssen2019} invoke the metallicity floor of GCs (i.e., the idea that there are few known GC with [Fe/H] $<-2$ to posit the properties (e.g., mass of galaxy and GC) and formation epoch of GCs that can survive a Hubble time. \eri\ and its GC are both lower in mass and metallicity than predicted by this models and formally fall into the `zone of avoidance' in which such a GC is predicted to be rare.  In this regime of predicted spare populations, \citet{kruijssen2019a} stress the importance of stochasticity, which means that although \eri\ is rare, it is fully consistent with these model predictions.

GCs have been considered as contributors to cosmic reionization \citep[e.g.,][]{renzini2017, boylan-kolchin2018}. For example, models presented in \citet{boylankolchin2018} suggest that even with modest values of birth-to-present day masses (i.e., $<5$) GCs in faint galaxies were likely to be significant contributors to the faint-end of the ultra-violet (UV) luminosity function ($z\sim4-10$).  Our characterization of \eri's GC is compatible with this scenario.  The formation epoch of the GC ($z\sim8$) is within the canonical reionization epoch, and the birth-to-present day mass ratio of 4 is also within range of these models.  The clear caveat is that \eri\ is only a single system and a key component of \citet{boylankolchin2018}'s models are a solid accounting of the GC number density, which our data do not help constrain.  

However, there is no consensus that GCs do contribute to reionization.  For example, GC formation models presented in \citet{choksi2018} and \citet{el-badry2019} predict that the bulk of GC formation took place after the Universe was reionized (i.e., $z\lesssim5$).  In these models, the GC in \eri\ is simply a rare occurrence at at high-z.

It is challenging to provide useful constraints on GC formation models from a single cluster in a single galaxy.  However, the combination of this type of archaeological analysis with the increased discovery rate of GCs in the local and early Universe \citep[e.g.,][]{carlsten2022, vanzella2022, pascale2023, vanzella2023, welch2023}, should help to increase empirical constraints on metal-poor GC formation across the galaxy mass spectrum and cosmic time.

\subsection{Present Day Mass and Mass-to-light Ratio}

\label{sec:masstolight}

Several studies exploit the fragility of the low-mass GC in \eri\ to constrain the shape of the inner dark matter profile of \eri.  Conclusions remain mixed; some studies suggest that a dark matter core is necessary for the long-term survival of \eri's GC \citep[e.g.,][]{amorisco2017, contenta2018, orkney2022}, others favor a cuspy dark matter profile \citep[e.g.,][]{zoutendijk2021}.

One key quantity in these models is the present day mass-to-light ratio of the GC.  From the CMD-based GC mass, we can estimate the present-day stellar mass by accounting for stellar evolution effects. The synthesis models of  \citet{conroy2009a} stellar evolution mass loss to be $\sim$40\%. Applying this to our CMD-based mass yields a present-day GC stellar mass of $M_{\star, z=0} \sim5.2\times10^3$ \msun.  At $z=0$, the GC is only $\sim4$\% of the galaxy's present day stellar mass, whereas at birth it was $\sim10$\%.  The latter is a more typical value for GCs in other galaxies at $z=0$ \citep[e.g.,][]{forbes2018}.

Using our present day stellar mass of the GC, we find  $M/L_V \sim 2.5$ for $M_V=-3.5$~mag \citep{crnojevic2016} and $M/L_V \sim 5$ for $M_V=-2.7$ \citep{simon2021}.  The former is within the scatter of the canonical mass-to-light ratios expected for an old \citep[i.e., $M/L_V = 2$, $\sigma\sim$1; e.g.,][]{conroy2009a, martin2008}, low-mass stellar population, whereas the latter is marginally high.  

As a check on the GC luminosity, we sum the fluxes of all GC stars within $2 r_h$ in F606W down to $m_{F606W} = 28$, which our ASTs indicate is 100\% complete.  We then transform F606W to V-band \citep{saha2011} and subtract the light contribution from 30\% of the GC stars that CMD modeling indicates are field star contaminants and the handful of MW foreground stars expected in the small \hst\ field of view.   CMD simulations indicate that unseen lower mass stars contribute 15\% more light than we measure.  In total, we find $M_{V, GC} \sim -3.4$ within $2 r_h$, in good agreement with \citet{crnojevic2016}, and yields a reasonable $M/L_V$.

For completeness, we also compute the mass-to-light ratio of the field population. Assuming that 40\% of the CMD-based stellar mass of the galaxy is lost to stellar evolution effects, we find a present day stellar mass $M_{\star, z=0}\approx1.5\times10^5$ \msun\ and $M_{\star, z=0}/L_V \approx 2.5$ for a galaxy luminosity of $M_{V} = -7.1$ \citep{crnojevic2016}.

\subsection{Comparison to GCs in the Local Group}

The GC in \eri\ is one of the oldest, most metal-poor, and lowest-mass GCs known in nearby dwarf galaxies (e.g., \citealt{forbes2018, beasley2019, larsen2020, larsen2022}).  GCs in more luminous dwarfs such as Fornax, NGC~6822, NGC~147, NGC~185, and WLM, tend to be more massive, metal-rich, and/or younger. Peg~DIG is the lowest-mass ($M_{\star, z=0} \sim 10^7$ \msun) gas-rich dwarf galaxy to host a GC.  But the GC is younger, more metal-rich, and 10$\times$ more massive than \eri's GC \citep{cole2017}. A small number of faint GCs have been identified in dwarf galaxies of moderate luminosity. For example, \citet{cusano2016} report a putative faint GC ($M_{\rm V} = -4.6$) in And~XXV ($M_{\rm V} = -9.1$) and \citet{caldwell2017} re-discovered a faint GC ($M_{\rm V} = -3.5$; comparable to the GC in \eri) in And~I ($M_{\rm V} = -11.7$). The M31 galaxy and GC luminosities have been updated with the RR Lyrae distances from \citet{savino2022}.   

Even within more massive galaxies, only a handful of more metal-poor GCs are known. For example, \citet{larsen2020} report an M31 GC with $\langle$[Fe/H]$\rangle = -2.91\pm0.04$~dex, with a dynamical mass of $\sim10^6$ \msun.  Though comparable in metallicity to \eri's GC, this M31 GC is $\sim100$ times more massive than \eri's and is more luminous than the entirety of \eri\ as a galaxy, suggesting it originated in a different environment (i.e., this single GC may have more stellar mass than all of \eri\ as a galaxy).

More recently, \citet{martin2022} identified a disrupted GC in the MW with $\langle$[Fe/H]$\rangle = -3.38\pm0.2$~dex with a stellar mass of at least $M_{\star, z=0} \sim 8\times10^3$\msun.  This is more metal-poor than \eri's GC, but may have a comparable stellar mass, though \citet{martin2022} note severval challenges in accurately measuring the mass of a disrupted object.  Regardless, the mass and metallicity suggest this disrupted GC could have originated in an accreted dwarf galaxy.  However, backing out the progenitor characteristics of a faint substructures in the MW halo is extremely challenging, even with full phase space information available \citep[e.g.,][]{brauer2022}, highlighting the particular importance of \eri\ for connecting metal-poor GCs to their host galaxies.

\subsection{\eri\ and its GC in the high-redshift Universe}

The sensitivity and resolving power of \hst\ and the \textit{James Webb Space Telescope} have the potential to directly capture the birth of GCs in the early Universe.  Recently, several candidate proto-GCs at high redshift have been recently reported in the literature.  For example, \citet{vanzella2019} use gravitational lensing to identify a putative proto-GC forming \textit{in situ} in a dwarf galaxy at $z\sim6$.  The proto-GC is reported to have $r_h\lesssim13$~pc, $M_{\star} \sim 10^6$ \msun, $\Sigma_{\rm SFR} \gtrsim 10^{2.5}$ \msun\ yr$^{-1}$ kpc$^{-2}$ and $M_{\rm UV}(z\sim6) \sim -15.6$.  The host is reported as dwarf galaxy with $r_h\sim450$~pc, $M_{\star} \sim 4\times10^8$ \msun, $\Sigma_{\rm SFR} \gtrsim 10^{1.5}$ yr$^{-1}$ kpc$^{-2}$ and $M_{\rm UV}(z\sim6) \sim -17.3$.  In many cases, these quantities reflects best estimates given the coarse constraints of the data (objects are spatially unresolved meaning clusters and galaxies can be hard to distinguish; e.g., \citealt{bouwens2022}).

For comparison, we can make estimates for the high-z properties of \eri\ and its GC. For the field population, our CMD modeling (plus fact of $\sim2$ correction for surveyed area) yields a galaxy-wide SFR of  $\dot{M}_{\star}\sim 1.3\pm0.5\times10^{-3}$ \msun\ yr$^{-1}$, averaged over our time resolution of 300~Myr.  We cannot rule out a shorter duration of star formation, which would increase $\dot{M}_{\star}$.  For example,  \citet{gallart2021} posit that the duration of star formation could have been as short as 100~Myr, while analytic chemical evolution models suggest an \eri-like MDF could be the result of Type II enrichment in as a little at $\sim50$~Myr \citep[e.g.,][]{weinberg2017}. These limits suggest a range of SFRs from $\dot{M}_{\star}\sim 1-7 \times10^{-3}$ \msun\ yr$^{-1}$ which are similar to what is found in star-forming dwarf galaxies in the very local Universe at the present day \citep[e.g.,][]{lee2011, johnson2013}.  Cosmological simulations suggest galaxies in the same mass range as \eri\ should have very bursty SFHs with rapid fluctuations in SF on the order of $10-20$~Myr \citep[e.g.,][]{stinson2007, governato2010, el-badry2016, kimm2016, fitts2017, wheeler2019, applebaum2021, sameie2022}.  Unfortunately, our data cannot resolve such fine detail in the SFHs.

These same cosmological simulations suggest that star formation is generally concentrated in the central regions of low-mass galaxies at early times.  Motivated by these simulations, we assume that star formation occurred over a region 50-500~pc in radius \citep[e.g.,][]{fitts2017, applebaum2021}, from which we estimate an average star formation rate surface density of $\Sigma_{\rm SFR} \sim 0.001-0.9$ \msun\ yr$^{-1}$ kpc$^{-2}$. 

To facilitate further comparison to high-z observations of proto-GC candidates, we reconstruct the rest frame UV luminosity of \eri's field population using its SFH and well-established techniques from the literature \citep[e.g.,][]{weisz2014c, boylan-kolchin2015}. During its initial epoch of star formation, we find $M_{\rm UV} \sim -9\pm2.5$, assuming SF durations ranging from $50-300$~Myr with stochastic bursts of $\sim20$~Myr in duration.

We can make similar estimate for UV luminosity of the GC following \citet{boylan-kolchin2018a}. A formation duration of 1-5~Myr for the GC yields an $\dot{M \star}\sim1-5\times10^{-3}$ \msun\ yr$^{-1}$.  A GC birth radius of 1-10~pc gives $\Sigma_{\rm SFR} \sim 5-500$ \msun\ yr$^{-1}$ kpc$^{-2}$. Finally, assuming the GC is an single stellar population, the restframe UV luminosity of the GC at birth was $M_{\rm UV} \sim -12.5$~mag, which faded to $M_{\rm UV} \sim -8.5$~mag 100~Myr after it formed.

\eri\ and its GC are fainter than what can be directly observed in the early Universe.  For example, they are orders of magnitude less massive and luminous than what \citet{vanzella2019}, which is likely observing a MW progenitor \citep[e.g.,][]{boylan-kolchin2015}, where \eri\ is clearly a dwarf galaxy at all redshifts.  Through lensing with magnifications of $\mu \lesssim 30$, \citet{bouwens2022} reports the detection of objects as faint as $M_{UV} \sim -13$ at $z\sim6-8$, which approaches the maximum UV luminosity of the GC in \eri.

At birth, the GC was $\sim9$\% of the stellar mass of the galaxy.  If the cluster formed before the majority of the field population, as often appears the case in the recent simulations of \citet{sameie2022}, then it may have a higher fractional mass.  Our data cannot rule out the possibility of non-concurrent GC an galaxy formation.  On the other hand, if the amount of dynamical mass loss was lower than our estimate, the cluster birth mass would be lower, and it would constitute a lower fraction of the system's total stellar mass.  The $\sim10$\% ratio of GC-to-galaxy stellar mass at birth is similar to what is known in the local Universe, i.e., present day mass ratios \citep[e.g.,][]{forbes2018}.

In a broader context, \eri\ appears to be a fairly common type of high-z galaxy based on its location on the high-z UV luminosity function (UVLF; e.g., \citealt{atek2018}, \citealt{bouwens2022}).  Given its modest luminosity, it likely resides near the peak of the UVLF, assuming the high-z galaxy UVLF turns over as many studies argue it should \citep[e.g.,][]{oshea2015, gnedin2016, yung2019}. The exact luminosity of the putative turnover is not yet known as it appears to be fainter than the limits of the deepest \hst\ imaging \citep[e.g.,][]{atek2018, bouwens2022}.

\subsection{The Age of \eri\ in a cosmological context}

While our archaeological approach to characterizing GCs provides access to low-mass galaxy and clusters not detectable virutally anywhere else in the Universe, there are some important caveats.  First, even with the exquisitely deep \hst\ imaging, our time resolving power is limited to a few hundred Myr. Thus, we cannot discern between a short formation epoch (i.e., everything formed within $\sim10$~Myr) vs a more protracted episode.  This results in large uncertainties on certain quantities of interest such as $\Sigma_{\rm SFR}$ (Table \ref{tab:gc}).  Higher S/N imaging would improve the time resolution, but ultimately, the increased time resolution is limited by the small number of stars on the CMD.  Incorporating other information (e.g., external knowledge of the metallicity distribution function; \citealt{de-boer2012}, \citealt{dolphin2016}) should also improve the time resolution of the SFH and cluster age.

A second limitation is due to uncertainties in the stellar models themselves \citep[e.g.,][]{weisz2011a, dolphin2013}.  Though the multiple models we tested in this paper yield consistent ages, even slight differences of a few hundred Myr can translate to large uncertainties in redshift space.  Though our knowledge of stellar physics continues to improve, model systematics remain important to precisely mapping stellar ages to redshifts.

Finally, note that interpreting the age of \eri, along with other UFDs in the LG is complicated in light of uncertainties in our knowledge of cosmology due to the Hubble tension.  For example, new physics solutions, such as some variations of early dark energy (EDE) result in Universe that is substantially younger ($\sim12.8$~Gyr; e.g., \citealt{smith2022}) compared to its age from vanilla $\Lambda$CDM and a \textit{Planck} cosmology ($\sim13.8$~Gyr; e.g., \citealt{planck2020}).  However, \eri's formation epoch ($\sim13.5$~Gyr) is older than the age of an EDE Universe. Though multiple stellar models (and CMD fitting techniques) find consistently old age for \eri, additional uncertainties in stellar physics (e.g., convection) may also need to be considered determining absolute ages \citep[e.g.,][]{chaboyer1996a, chaboyer2017, omalley2017, joyce2022}.  Alternatively, since virtually all known UFDs around the MW appear to be older than $\sim13$~Gyr \citep[e.g.,][]{brown2014, weisz2014a, sacchi2021}, it may indicate that physics-based solutions to the Hubble tension resulting in ages younger than $\sim$13~Gyr may not be consistent with all available measurements of the Universe's age.  Similarly, \citet{boylan-kolchin2021} note that uncertainties in the cosmological model due to the Hubble tension introduces a minimum uncertainty of 5\% in the age-redshift mapping, which sets a floor on any comparison between ages of stars and galaxies and ages inferred from redshifts in the early Universe.

\begin{acknowledgments}
We thank the anonymous referee for a constructive and thorough review. The authors thank Nick Choksi, Mike Boylan-Kolchin, Kareem El-Badry, and Diederik Kruijssen for helpful comments on early drafts.  Support for this work was provided by NASA through grants HST-GO-13768, HST-GO-15746, HST-GO-15901, HST-GO-15902, and HST-AR-16159 from the Space Telescope Science Institute, which is operated by AURA, Inc., under NASA contract NAS5-26555. This research has made use of NASA’s Astrophysics Data System Bibliographic Services
\end{acknowledgments}

\vspace{5mm}
\facilities{HST(ACS)}

\software{ This research made use of routines and modules from the following software packages: \texttt{Astropy} \citep{Astropy}, \texttt{DOLPHOT} \citep{dolphin2016}, \texttt{IPython} \citep{IPython}, \texttt{Matplotlib} \citep{Matplotlib}, \texttt{NumPy} \citep{Numpy} and \texttt{SciPy} \citep{Scipy}
          }

\bibliography{main_bib, citations}
\bibliographystyle{aasjournal}

\end{document}